\documentstyle[12pt]{article}
\newlength{\pubnumber} 

\catcode`\@=11
\@addtoreset{equation}{section}
\def\section{\@startsection{section}{1}{\z@}{3.5ex plus 1ex minus .2ex}
 {2.3ex plus .2ex}{\large\bf}}
\def\subsection{\@startsection{subsection}{2}{\z@}{2.3ex plus .2ex}
 {2.3ex plus .2ex}{\bf}}

\begin{document}

\begin{titlepage}
\samepage{
\setcounter{page}{1}
\rightline{IASSNS-HEP-95/24}
\rightline{\tt hep-th/9505018}
\rightline{published in {\it Phys.\ Rev.\ Lett.}\/ {\bf 75} (1995) 2646}
\rightline{April 1995}
\vfill
\begin{center}
 {\Large \bf Making Ends Meet:\\
     String Unification and Low-Energy Data\\}
\vfill
 {\large Keith R. Dienes\footnote{
   E-mail address: dienes@sns.ias.edu}
   $\,$and$\,$ Alon E. Faraggi\footnote{
   E-mail address: faraggi@sns.ias.edu}\\}
\vspace{.12in}
 {\it  School of Natural Sciences, Institute for Advanced Study\\
  Olden Lane, Princeton, N.J.~~08540~ USA\\}
\end{center}
\vfill
\begin{abstract}
  {\rm
  A long-standing problem in string phenomenology has been the fact
  that the string unification scale disagrees
   with the
  GUT scale obtained by extrapolating low-energy data
  within the framework of the minimal supersymmetric standard model (MSSM).
  In this paper we examine several effects that may modify the minimal
  string predictions and thereby bring string-scale unification into agreement
  with
  low-energy data.  These include heavy string threshold corrections,
  non-standard hypercharge normalizations, light SUSY thresholds, intermediate
  gauge structure, and thresholds arising from extra matter beyond the
  MSSM.
  We explicitly evaluate these contributions within a variety of realistic
  free-fermionic string models, including the flipped
   $SU(5)$, $SO(6)\times SO(4)$,
  and various $SU(3)\times SU(2)\times U(1)$ models, and find
  that most of these sources do not substantially alter the minimal
  string predictions.
  Indeed, we find that the only way to
   reconcile string unification with low-energy data is through
  certain types of extra matter.
  Remarkably, however, many of the realistic string models
  contain precisely this required matter in their low-energy spectra.
  }
\end{abstract}
\vfill
\smallskip}
\end{titlepage}

\setcounter{footnote}{0}

\def\beq{\begin{equation}}
\def\eeq{\end{equation}}
\def\beqn{\begin{eqnarray}}
\def\eeqn{\end{eqnarray}}
\hyphenation{su-per-sym-met-ric non-su-per-sym-met-ric}
\hyphenation{space-time-super-sym-met-ric}
\hyphenation{mod-u-lar mod-u-lar--in-var-i-ant}


\setcounter{footnote}{0}

Heterotic string theories have a number of properties which
make them the leading candidates for a unified theory of the fundamental
particles and interactions.  Not only do they
provide a first-quantized description of gravity, for example,
but they also incorporate $N=1$ supersymmetric gauge
theories as their low-energy limits.
This is an important feature, for such supersymmetric gauge theories
are natural extensions of the Standard Model which are in
agreement with all low-energy experimental data.
Moreover, one such theory in particular, namely the minimal supersymmetric
Standard Model (MSSM), provides a successful scenario for gauge coupling
unification,
with a predicted unification scale $M_{\rm MSSM}\approx 2\times 10^{16}$ GeV.
String theories, by contrast, predict gauge coupling unification
at a somewhat larger unification scale,
typically $M_{\rm string}\approx g_{\rm string}\times 5\times 10^{17}$ GeV
where $g_{\rm string}\approx 0.8$ at unification.
This discrepancy between these two unification scales
implies that string theory
naively predicts incorrect values for
the electroweak and strong couplings
$\sin^2\theta_W$ and
$\alpha_{\rm strong}$ at the $Z$ scale.
Resolving this discrepancy
and ``making the two ends meet'' is therefore one of the major problems
confronting string phenomenology.

There are many possible effects which may account for this
discrepancy and alter the running of the gauge couplings
between the high and low energy scales.
First, there are the so-called ``heavy
string threshold corrections'' which represent the contributions
from the massive Planck-scale string states
that are otherwise neglected in an analysis of
the purely low-energy ({\it i.e.}, massless) string spectrum.
Second, there are potential corrections due to the fact
that in string theory, the normalization of the $U(1)$ hypercharge
need not take the standard value that it has in various field-theoretic
GUT models.
Third, if supersymmetry is ultimately broken
at the TeV-scale, the required SUSY-breaking terms
will lead to additional light SUSY thresholds.
Fourth, in various GUT scenarios, there can be corrections due to
the presence of non-trivial gauge structure at intermediate
scales.
Finally, there may also be contributions from additional
exotic states beyond those predicted by the MSSM.
While such states are not expected in standard field-theoretic GUT
scenarios, we shall see that
they appear naturally in certain self-consistent string models.

Many of these effects have been discussed previously,
each in an abstract setting and in isolation.
However, within the tight constraints of a given realistic
string model, the mechanisms giving rise to three
generations and the MSSM gauge group
may prove inconsistent with, for example,
large threshold corrections or extra non-MSSM matter.
Moreover, the increased complexity of the known realistic string models
may substantially alter previous expectations
based on simplified or idealized scenarios.
It is therefore important to rigorously calculate all
of these effects simultaneously,
within the context of a wide variety of actual
realistic string models,
in order to determine which path to unification (if any)
such models actually take.

In this paper, we shall present the results of the
first such systematic evaluation.
A more complete discussion of our analysis and results will be presented in
Ref.~\cite{bigpaper}.
Surprisingly, we find that
most of these effects in these models cannot resolve the
discrepancy
between string unification and low-energy data.
In particular, they do not modify the one-loop renormalization
group equations (RGE's) in a manner that would yield the correct values
of $\alpha_{\rm strong}$ and $\sin^2 \theta_W$ at the $Z$-scale
when starting at the string scale.
Indeed, we find that only the effect of certain extra
exotic states has the potential to bring the low-energy data
into agreement with string unification.
Remarkably, these states, which take the form of extra color triplets and
electroweak
doublets with special $U(1)$ quantum numbers, appear naturally in
the realistic free-fermionic string models.
Thus string theory, which predicts an unexpectedly high
unification scale $M_{\rm string}$, in many cases also simultaneously
predicts precisely the extra exotic particles needed to reconcile
this higher scale with low-energy data.

The string models that we have chosen for our analysis
include the flipped $SU(5)$ model of Ref.~\cite{flipped},
the $SO(6)\times SO(4)$ model of Ref.~\cite{alrmodel},
and various string models \cite{aefone,aeftwo}
in which the Standard-Model gauge group
$SU(3)\times SU(2)\times U(1)$
is realized directly at the Planck scale.
As required, all of these models have $N=1$ spacetime
supersymmetry, and contain exactly three generations in
their massless spectra.
Moreover, these models also naturally incorporate the fermion
mass hierarchy with a heavy top-quark mass.
These models can all be realized in the free-fermionic
construction \cite{freefermions}, and their phenomenological
properties are ultimately a consequence of their common
underlying $Z_2\times Z_2$ orbifold structure \cite{orbifold}.  In the
free-fermionic construction, this structure is realized
through the so-called ``NAHE set''
of boundary-condition basis vectors.

String theory predicts in general that at tree level,
the gauge couplings $g_i$ corresponding to each gauge group
factor $G_i$ realized at Ka\v{c}-Moody level $k_i$
will unify with the gravitational coupling constant $G_N$
according to \cite{Ginsparg}
\beq
     g^2_{\rm string}~=~8\pi \,{G_N\over\alpha'}~=~ g_i^2\,k_i
       ~~~~{\rm for~all}~i~
\label{treelevelrelation}
\eeq
where $\alpha'$ is the Regge slope.
At the one-loop level, however, these relations are modified to
\beq
     {{16\pi^2}\over{g_i^2(\mu)}}~=~k_i{{16\pi^2}\over{g_{\rm string}^2}}~+~
     b_i\,\ln{ M^2_{\rm string}\over\mu^2}~+~\Delta_i^{\rm (total)}
\label{onelooprunning}
\eeq
where $b_i$ are the one-loop beta-function coefficients, and where
the $\Delta_i^{\rm (total)}$
represent the combined corrections from each of the effects discussed above.
Thus, in a given realistic string model,
we can use Eq.~(\ref{onelooprunning})
to find the expected values of the
strong and electroweak gauge couplings at the $Z$-scale,
and thereby obtain explicit expressions for $\alpha_{\rm strong}(M_Z)$ and
$\sin^2\theta_W(M_Z)$.

We organize the calculation as follows.
In each case we take $\alpha_{\rm e.m.}(M_Z)=1/127.9$
as a fixed input parameter, and leave $k_1$ arbitrary.
We then solve Eqs.~(\ref{onelooprunning}) for $i=1,2,3$ simultaneously
in order to eliminate the direct dependence
on $g_{\rm string}$ from the first term on the right
sides of Eqs.~(\ref{onelooprunning}).  A small dependence
on $g_{\rm string}$ remains through $M_{\rm string}$,
and in all subsequent numerical calculations we will
allow $M_{\rm string}$ to vary in the range $(2-7)\times 10^{17}$
GeV in order to account for this.
In each case we initially assume the MSSM spectrum
between the Planck scale and the $Z$-scale,
and treat all perturbations of this scenario through
effective correction terms.
The expression for $\alpha_{\rm strong}(M_Z)$
then takes the general form
\beqn
    \alpha^{-1}_{\rm strong}(M_Z)&=&
      \Delta_{\rm MSSM}^{(\alpha)} +
      \Delta_{\rm h.s.}^{(\alpha)} +
      \Delta_{\rm l.s.}^{(\alpha)} \nonumber\\
      &&~+ \Delta_{\rm i.g.}^{(\alpha)} +
      \Delta_{\rm i.m.}^{(\alpha)} +
      \Delta_{\rm other}^{(\alpha)},
\label{genform}
\eeqn
and likewise for $\sin^2\theta_W(M_Z)$ with corresponding
corrections $\Delta^{\rm (sin)}$.
Here $\Delta_{\rm MSSM}$ represents the one-loop
contributions from the MSSM spectrum alone, and
the remaining
$\Delta$ terms respectively correspond
to heavy string thresholds, light SUSY thresholds,
intermediate-scale gauge structure, and
intermediate-scale extra matter.
Each of these $\Delta$ terms has an algebraic
expression in terms of $\alpha_{\rm e.m.}$ as well
as model-specific parameters such as $k_1$,
the beta-function coefficients, and any
appropriate intermediate mass scales.
Finally, we also include in our analysis
various two-loop and Yukawa-coupling effects,
as well as explicit correction factors arising from scheme conversion
(from the SUSY-based $\overline{DR}$ scheme to the
$\overline{MS}$ scheme relevant for comparisons
with low-energy data).
These are represented by $\Delta_{\rm other}$.

We have explicitly calculated each of these $\Delta$
contributions for each of the realistic string models
discussed above.
Our results are as follows.

 \underbar{\sl Two-loop, Yukawa, and scheme conversion:}~~
We first focus on these ``other'' corrections.
As discussed above, these are calculated assuming the MSSM spectrum
between the string unification and $Z$ scales.
To estimate the size of the second-loop
corrections, we run the one- and two-loop RGE's
for the gauge couplings and take the difference.
Likewise, to estimate the Yukawa-coupling corrections, we
evolve the two-loop RGE's for the gauge couplings coupled
with the one-loop RGE's for the heaviest-generation Yukawa couplings,
assuming $\lambda_t\approx 1$ and $\lambda_b=\lambda_\tau\approx 1/8$
at the string unification scale.
We then subtract the two-loop non-coupled result.
As indicated above,
these differences are then each averaged for different values
of $M_{\rm string}$.
Including the standard scheme-conversion corrections as well,
we find
\beq
     \Delta_{\rm other}^{(\sin)}\approx 6.13\times 10^{-3} ~,~~~~~
     \Delta_{\rm other}^{(\alpha)}\approx 0.700     ~.
\eeq

 \underbar{\sl Non-standard hypercharge normalizations:}~~
We have studied the effect of varying the value of $k_1$ within
the RGE's, as proposed, for example, in Ref.~\cite{ibanez}.
Although $k_1=5/3$ in standard $SO(10)$ GUT
scenarios, in string theory the hypercharge normalizations
can generally be different.  We find, however, that
the experimental discrepancies are resolved only for
values $k_1\approx 1.4$, whereas each of these
string models ultimately turns out to have $k_1\geq 5/3$.
Thus, this effect cannot resolve the discrepancy between
the GUT and string scales in these models.
Moreover, arguments exist \cite{dfm} showing that
$k_1\geq 5/3$ in any free-field
heterotic string model containing the MSSM
gauge group and spectrum
realized at level $k_2=k_3=1$.
This constraint arises because $k_1$
is related to the embedding of the hypercharge in terms
of simple worldsheet $U(1)$ currents with
proper conformal-field-theory normalizations.
One then finds \cite{dfm} that
the minimal embedding yielding the correct hypercharges
for MSSM particles corresponds to $k_1=5/3$.
Thus, it is not likely that other level-one
models can exploit this effect either.

 \underbar{\sl Heavy string threshold corrections:}~~
Calculating the heavy string threshold corrections within
the realistic free-fermionic models is by far the most complex part of
the analysis, for there are numerous subtleties arising due
to twisted fermionic boundary conditions and
highly non-trivial free-fermionic realizations of the gauge groups
due to occasional enhanced gauge symmetries.
A full account of our analysis will be presented in Ref.~\cite{bigpaper}.
We follow Kaplunovsky \cite{kaplunovsky} in defining
the threshold corrections $\Delta_i$ corresponding to
each gauge-group factor $G_i$ as
\beq
   \Delta_{i}~\equiv~ \int_{\cal F} {d^2\tau\over {\rm Im}\,\tau }
      \left\lbrace
          {\rm Tr}\, \overline{Q}_h^2 Q_i^2\,q^H \overline{q}^{\overline{H}}
          ~-~ b_i\right\rbrace
\label{deltadef}
\eeq
where $\overline{Q}_h$ is the spacetime helicity operator,
$Q_i$ is the internal gauge charge operator for the group $G_i$,
$b_i$ is the corresponding one-loop beta-function coefficient,
and $H$ and $\overline{H}$ are respectively the
left- and right-moving worldsheet Hamiltonians.
The trace is then performed over all sectors in the theory and
all mass levels in each sector, and the result with $b_i$ subtracted
is then integrated over the fundamental domain ${\cal F}$ of the
modular group.
In the realistic models we consider, there are typically thousands
of sectors which make non-vanishing contributions to $\Delta_i$,
and great care has to be taken with regards to the GSO projection
phases in order to verify that the desired spectrum is produced.
Other subtleties include deriving explicit expressions for $Q_i^2$
in terms of the individual charges $Q_L$ corresponding to each
worldsheet fermion;  in the cases of string models with enhanced
gauge symmetries, this embedding of the gauge charges within the free
fermions can be highly non-trivial.
Finally, and equally importantly, one must perform the integrations
in Eq.~(\ref{deltadef}) in such a way that numerical errors
are minimized and all logarithmic divergences cancelled.
Details can be found in Ref.~\cite{bigpaper}.
Note that because these are $(2,0)$ models [rather than
$(2,2)$], one cannot use various moduli-dependent
approximations found in the literature.

The various quantities $\Delta_{i}$ one obtains must then
be appropriately combined in order to yield expressions for
$\Delta_{\rm h.s.}^{(\alpha,{\rm sin})}$ in Eq.~(\ref{genform}).
In particular, we find
\beqn
  \Delta_{\rm h.s.}^{(\sin)} &=& {1\over 2\pi}\, {k_1 \over{k_1+1}}\,
       \alpha_{\rm e.m.}\, {\Delta_2-\Delta_{\hat Y}\over 2} \nonumber\\
    \Delta_{\rm h.s.}^{(\alpha)}&=&
         {-1\over 4\pi}\,{1\over k_1+1}\,\left\lbrack
           k_1(\Delta_{\hat Y}-\Delta_3) + (\Delta_2 -\Delta_3)\right\rbrack~
\eeqn
where $\Delta_{3,2,\hat Y}$ respectively correspond to the strong, electroweak,
and properly normalized weak hypercharge groups.
It is important that only the relative {\it differences}\/ of these quantities
appear, for the definition in Eq.~(\ref{deltadef}) neglects certain
group-independent additive factors.

Our results are then as follows.
For the $SU(3)\times SU(2)\times U(1)_{\hat Y}$ model of Ref.~\cite{aefone},
we find
\beq
     \Delta_{\hat Y}-\Delta_{3}=  5.0483~,  ~~~~~~~
     \Delta_{\hat Y}-\Delta_{2}=  1.6137~,
\label{DELTAS278}
\eeq
so that $\Delta_{\rm h.s.}^{(\sin)}= -6\times 10^{-4}$
and $\Delta_{\rm h.s.}^{(\alpha)}= -0.3536$.
Since the hypercharge operator in this model has
the standard $SO(10)$ embedding, we have taken $k_1=5/3$.
Unfortunately, we see from the signs and sizes of these results
that they effectively {\it increase}\/ the
string unification scale slightly, and thereby
enhance the disagreement with experiment in this model.

In the remaining models, the Standard Model
gauge group is realized only after the Planck-scale
gauge group is broken at an intermediate scale $M_I$.
We shall analyze the effects of such intermediate
scales below, and thus assume for the purposes of this analysis
that $M_I\approx M_{\rm string}$.
For the flipped $SU(5)\times U(1)$ model of Ref.~\cite{flipped},
we then find $\Delta_{1}-\Delta_5 = 7.681$,
so that after combining these results in a manner
appropriate for the flipped SU(5) symmetry-breaking
scenario [which for $M_I=M_{\rm string}$ amounts
to taking $\Delta_2=\Delta_3=\Delta_5$ and
$\Delta_{\hat Y}=(\Delta_5 +24\Delta_1)/25$], we obtain
the final low-energy corrections
$\Delta_{\rm h.s.}^{(\sin)}= -2\times 10^{-3}$
and $\Delta_{\rm h.s.}^{(\alpha)}=-0.29$.
This again implies a slight effective increase in
the string unification scale.\footnote{
      Note that our result for
     $\Delta_1-\Delta_5$ in this model disagrees by a factor
     of approximately three  with that
     obtained in Ref.~\cite{anton}.
      The size of our result, however, is more in line
     with those from the other realistic models we examine,
     as well as from previous heavy threshold calculations
     in various orbifold and Type-II models \cite{others}.  }
Likewise, for the $SO(6)\times SO(4)\simeq SU(4)\times
SU(2)_L\times SU(2)_R$ model of Ref.~\cite{alrmodel},
we find
\beq
      \Delta_{2_L}-\Delta_{4}= 8.4763~, ~~~
      \Delta_{2_R}-\Delta_{4}= 2.0483~,
\label{DELTASO64}
\eeq
so that, after the Pati-Salam symmetry-breaking
scenario [which for $M_I=M_{\rm string}$ implies that
we take $\Delta_3=\Delta_4$,
$\Delta_2=\Delta_{2_L}$,
and $\Delta_{\hat Y}=(2\Delta_4+3\Delta_{2_R})/5$],
we obtain the low-energy results
$\Delta_{\rm h.s.}^{(\sin)}= -2.8\times 10^{-3}$
and $\Delta_{\rm h.s.}^{(\alpha)}=-0.3141$.
This too implies a slight effective increase in the
string unification scale.
Finally, for the $SU(3)\times SU(2)\times U(1)$
model of Ref.~\cite{aeftwo},
we similarly find after a complicated pattern of
symmetry-breaking that
$\Delta_{\rm h.s.}^{(\sin)}= -1.2\times 10^{-3}$
and $\Delta_{\rm h.s.}^{(\alpha)}=-0.335$,
again implying that $M_{\rm string}$
is effectively increased.

It is an important observation that the sizes of
the threshold corrections are very small in
all of these realistic string models, and thus
do not greatly affect (either positively or negatively)
the magnitude of the string unification scale.
Thus, we conclude that these threshold corrections cannot by
themselves resolve the experimental discrepancy.
Moreover,
despite the fact that such threshold
corrections receive contributions
from infinite towers of massive string states,
we have been able to provide a general model-independent
argument \cite{bigpaper}
which explains why these corrections
must always be naturally suppressed in string theory
(except of course for large moduli).
This suppression arises due to the modular properties of
the integrand of Eq.~(\ref{deltadef}), in particular the
insertion of the charge and helicity operators into the
trace.  These operators increase the modular weight of the
integrand, and thereby suppress the
value of the integral in tachyon-free models.
Further details can be found in Ref.~\cite{bigpaper}.
Hence alternative string models are not likely to have
significantly larger threshold corrections.

 \underbar{\sl Intermediate gauge structure:}~~
 In the above calculations, we have implicitly assumed
that $M_I\approx M_{\rm string}$.  In general,
however, we will have $M_I\leq M_{\rm string}$, and this
has the potential to change the analysis.  We have
investigated this possibility, but find that
in the above models,
taking values $M_I< M_{\rm string}$ surprisingly
only {\it enhances}\/ the disagreement with experiment.
Thus, once again, this cannot eliminate the experimental
discrepancy.

 \underbar{\sl Light SUSY thresholds:}~~
 As a perturbation on the assumption of the MSSM spectrum
from the string scale to the $Z$ scale, one can also
consider the effects of the light SUSY thresholds
that arise from SUSY-breaking.  These effects can
ultimately be parametrized in terms of the four soft SUSY-breaking
parameters $\lbrace m_0, m_{1/2}, m_h , m_{\tilde h}\rbrace$,
assuming either universal or non-universal boundary terms
for the sparticle masses.  We find, however, that even in the ``best-case''
scenario, these effects cannot resolve the discrepancy
with low-energy data.  In Fig.~1(a) we have plotted the
values of $\lbrace \sin^2\theta_W(M_Z),\alpha_{\rm strong}(M_Z)\rbrace$
obtainable under variations of $M_{\rm string}$ and
these SUSY-breaking parameters, assuming universal boundary conditions.
It is clear that the experimentally measured values of
$\sin^2 \theta_W(M_Z)\approx 0.231$ and
$\alpha_{\rm strong}(M_Z)\approx 0.125$
are not reached.

\input epsf
\begin{figure}[t]
\centerline{\epsfxsize 4.0 truein \epsfbox {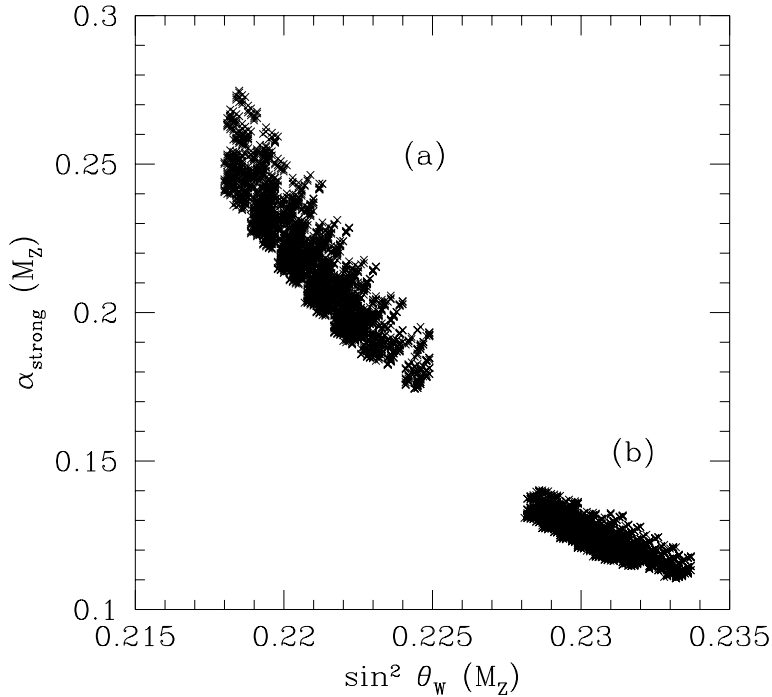}}
\caption{~~Scatter plot of $\lbrace\sin^2\theta_W(M_Z),
       \alpha_{\rm strong}(M_Z)\rbrace$
       for various values of $\lbrace m_0,m_{1/2},\break
        m_h,m_{\tilde h},M_{\rm string}\rbrace$. Region (a) assumes
         the MSSM spectrum as discussed in the text, while (b)
       also includes the effects of the string-predicted
         extra matter.}
\label{figureone}
\end{figure}

 \underbar{\sl Extra string-predicted matter:}~~
Finally we consider the effects of extra string-predicted
matter beyond the MSSM \cite{aeftwo,Gaillard}.
While the introduction of such matter may seem {\it ad hoc}\/ from the
low-energy point of view,
our analysis thus far indicates that this is the {\it only}\/ way
by which the experimental discrepancies might be reconciled
in realistic free-fermionic models.
Furthermore,
such matter appears naturally in
these models, and its effects
must be included.
This matter typically takes the form
of additional color triplets and electroweak doublets, in vector-like
representations.  Note that
the masses and $U(1)$ quantum numbers of such
states are highly model-dependent.
The masses, in particular,
can be calculated from cubic or higher
non-renormalizable terms in the superpotential,
and depend on the particular SUSY-breaking
scheme employed.  In general, however, these
masses are often much smaller than the string scale.
For example, in one model,
the mass scale of such an additional color
triplet was estimated \cite{massestimate} to be of the order
of $10^{11}$ GeV.

We have analyzed the effects that such matter can have on the
unification of the couplings.  Despite the presence of this
new intermediate scale, however, we find that successful low-energy
predictions are still generically difficult to obtain.
Indeed, in order to accommodate the low-energy data, we find on general grounds
that
extra triplets {\it and}\/ doublets must be present simultaneously,
and moreover that this extra matter must have particular
hypercharge assignments so as
to modify the running of the strong and electroweak couplings
without substantially affecting the $U(1)$ coupling.
These include, for example, $SU(3)\times SU(2)$ representations such
as $(3,2)_{1/6}$,
$(\overline{3},1)_{1/3}$,
$(\overline{3},1)_{1/6}$,
and $(1,2)_0$, all of which have large values of $b_2$ and $b_3$,
but relatively small values of $b_1$.
Many of the realistic free-fermion models ({\it e.g.},
that in Ref.~\cite{aefone}) do not have such exotic
representations and can actually be ruled out on this basis.

Remarkably, however, some of the other realistic string models
predict extra matter with exactly the required hypercharge assignments
and in exactly the proper combinations for successful string-scale
unification.
For example, the model of Ref.~\cite{aeftwo}
contains in its spectrum
two pairs of $(\overline{3},1)_{1/3}$
color triplets with beta-function coefficients $(b_3,b_2,b_1)=(1/2,0,1/5)$,
one pair of $(\overline{3},1)_{1/6}$ triplets with $b_i=(1/2,0,1/20)$,
and three pairs of $(1,2)_{0}$ doublets with $b_i=(0,1/2,0)$.
Clearly, this exotic matter cannot be fit into the standard
$SO(10)$ representations.   Nevertheless, we find that this
particular combination of representations and hypercharge
assignments opens up a sizable window
in which the low-energy data and string unification can
be reconciled.  For example, we find that if these triplets
all have equal masses in the approximate range
$ 2\times 10^{11} \leq M_3 \leq 7\times 10^{13}$ GeV
with the doublet masses in the corresponding range
$ 9\times 10^{13} \leq M_2 \leq 7\times 10^{14}$ GeV,
then the discrepancy is removed.  This situation is
illustrated in Fig.~1(b), which plots the same points
as in Fig.~1(a) except that this extra matter is now included
in the analysis.
Details and other scenarios for each of the realistic
string models will be discussed in Ref.~\cite{bigpaper}.
Of course, we emphasize that it is still necessary to verify
that this extra matter can actually have the needed masses
in these models.

We conclude, then, that string-scale unification places tight
constraints on realistic free-fermion string models,
and that only the appearance of extra exotic matter in particular
representations can possibly resolve the experimental discrepancies.
It is of course an old idea that the presence of extra matter
can resolve the discrepancy between the GUT and string unification
scales.  What is highly non-trivial, however, is that this now appears
to be the {\it only}\/ way in which string theory can solve the
problem.  Furthermore, we have seen that a subset of these realistic
models actually manage to satisfy these constraints in precisely this manner,
naturally giving rise to the required sorts of extra states with the proper
non-standard hypercharges to do the job.  It will therefore be interesting
to see whether this string-predicted extra matter has masses in
the appropriate ranges, and to determine the effects that this
extra matter might have on low-energy physics.
Such work is in progress \cite{progress}.

\bigskip
\medskip
\leftline{\large\bf Acknowledgments}
\medskip

We thank I. Antoniadis, S. Chaudhuri, S.-W. Chung, L. Dolan, E. Kiritsis,
J. Louis, J. March-Russell, R. Myers, J. Pati, M. Peskin, F. Wilczek,
and E. Witten for discussions.
This work was supported in part by DOE Grant No.\ DE-FG-0290ER40542.

\bigskip
\bigskip

\vfill\eject

\bibliographystyle{unsrt}

\end{document}